\documentclass[fleqn]{article}
\usepackage{exscale} \usepackage{makeidx}
\usepackage{amsmath} \usepackage{amssymb}
\usepackage{amsfonts} \usepackage{amstext} \usepackage{amsbsy}
\usepackage[mathscr]{eucal} 

\title{Hamiltonians for a general dilaton gravity theory on a
spacetime with a non-orthogonal, timelike or spacelike outer
boundary} 
\author{M.~Cadoni,\,$^a$\,\thanks{E-mail address: \texttt{cadoni@ca.infn.it}}
\and
P.~G.~L.~Mana\,$^b$\,\thanks{E-mail address: \texttt{mana@ca.infn.it}}}

\date{\footnotesize
$^a$\emph{Dipartimento di Fisica, Universit\`a
di Cagliari,\\ Cittadella Universitaria, I-09042 Monserrato, Italy},\\
and\\
\emph{INFN, Sezione di Cagliari, Italy}.\\[2ex]
$^b$\emph{Dipartimento di Fisica, Universit\`a di
Cagliari,\\ Cittadella Universitaria, I-09042 Monserrato, Italy.}\\[1em]
September 2000\\[1em]
{INFNCA-TH0020}}

\allowdisplaybreaks[1]
\newcommand{\dens}[1]{\boldsymbol{#1}}
\newcommand{\inte}[1]{\mathfrak{#1}}
\newcommand{\oper}[1]{\boldsymbol{\mathbf{#1}}}
\newcommand{\dila}[1]{{#1^\dil}\vphantom{\bar{#1}}}
\newcommand{\vac}[1]{\smash{\underset{\smash{\bar{\vphantom{o}}}}{#1}}\vphantom{#1}}
\newcommand{\R}{\mathbb{R}}
\newcommand{\yd}{d}
\newcommand{\ym}{\mathcal{M}}
\newcommand{\ymo}{\vac{\ym}}
\newcommand{\yf}{\mathcal{\partial}}
\newcommand{\yb}{\mathcal{B}}
\newcommand{\ys}{\mathcal{S}}
\newcommand{\ysp}{\Tilde{\ys}}
\newcommand{\yi}{\mathcal{I}}
\newcommand{\yp}{\mathcal{P}}
\newcommand{\ep}{\varepsilon}
\newcommand{\dil}{\eta}
\newcommand{\dilo}{\vac{\dil}}
\newcommand{\yub}{{\Tilde{\yups}}}
\newcommand{\yubo}{{\vac{\yub}}}
\newcommand{\yus}{u}
\newcommand{\yups}{n}
\newcommand{\yupso}{{\vac{\yups}}}
\newcommand{\yupb}{{\Tilde{\yus}}}
\newcommand{\yns}{N}
\newcommand{\yvs}{V}
\newcommand{\yvso}{{\vac{\yvs}}}
\newcommand{\ynp}{\Tilde{\yns}}
\newcommand{\yvp}{\Tilde{\yvs}}
\newcommand{\ygb}{\gamma}
\newcommand{\ygs}{h}
\newcommand{\ygp}{\sigma}
\newcommand{\ygmo}{{\vac{g}}}
\newcommand{\ygme}{\sqrt{-g}}
\newcommand{\ygbe}{\sqrt{|\ygb|}}
\newcommand{\ygpe}{\sqrt{\ygp}}
\newcommand{\ygse}{\sqrt{\ygs}}
\newcommand{\ygmd}{\dens{\ygme}}
\newcommand{\ygbd}{\dens{\ygbe}}
\newcommand{\ygpd}{\dens{\ygpe}}
\newcommand{\ygsd}{\dens{\ygse}}
\newcommand{\ygsdo}{\vac{\ygsd}}
\newcommand{\ydm}{\boldsymbol{\nabla}}
\newcommand{\ydmo}{{\vac{\ydm}}}
\newcommand{\ydb}{\boldsymbol{\Delta}}
\newcommand{\yds}{\oper{D}}
\newcommand{\ypmsc}{\Xi}
\newcommand{\ypmdilsc}{\dila{\ypm}}
\newcommand{\ypbsc}{\Pi}
\newcommand{\ypbdilsc}{\dila{\ypb}}
\newcommand{\ypssc}{P}
\newcommand{\ypsdilsc}{\dila{\yps}}
\newcommand{\yppsc}{\pi}
\newcommand{\yppdilsc}{\dila{\ypp}}
\newcommand{\ypm}{\dens{\ypmsc}}
\newcommand{\ypb}{\dens{\ypbsc}}
\newcommand{\yps}{\dens{\ypssc}}
\newcommand{\ypso}{{\vac{\yps}}}
\newcommand{\ypp}{\dens{\yppsc}}
\newcommand{\ypmdil}{\dens{\ypmdilsc}}
\newcommand{\ypbdil}{\dens{\ypbdilsc}}
\newcommand{\ypbodil}{\vac{\ypbodil}}
\newcommand{\ypsdil}{\dens{\ypsdilsc}}
\newcommand{\yppdil}{\dens{\yppdilsc}}
\newcommand{\ycb}{\Theta}
\newcommand{\ycs}{K}
\newcommand{\ycpb}{{\Tilde{\ycps}}}
\newcommand{\ycpso}{\vac{\ycps}}
\newcommand{\ycps}{\theta}
\newcommand{\ycpbo}{\vac{\ycpb}}
\newcommand{\yrm}{R_{\ym}\vphantom{R}}
\newcommand{\yrs}{R_{\ys}\vphantom{R}}
\newcommand{\ypz}{\beta}
\newcommand{\ypa}{\alpha}
\newcommand{\ypl}{\lambda}
\newcommand{\yli}{\inte{I}}
\newcommand{\ylio}{\vac{\yli}}
\newcommand{\yhi}{\inte{H}}
\newcommand{\yhio}{\vac{\yhi}}
\newcommand{\yhii}{\Tilde{\yhi}}
\newcommand{\yhiio}{\vac{\yhii}}
\newcommand{\yhd}{\dens{H}^\perp}
\newcommand{\ymd}{\dens{H}}
\newcommand{\yim}{\mu}
\newcommand{\yjm}{\nu}
\newcommand{\ykm}{\tau}
\newcommand{\ylm}{\sigma}
\newcommand{\yib}{\yim}
\newcommand{\yjb}{\yjm}
\newcommand{\yis}{\yim}
\newcommand{\yjs}{\yjm}
\newcommand{\yks}{\ykm}
\newcommand{\yip}{\yim}
\newcommand{\yjp}{\yjm}
\newcommand{\fdil}{f}
\newcommand{\fdilo}{\vac{\fdil}}
\newcommand{\kdil}{k}
\newcommand{\pdil}{p}
\newcommand{\var}{\delta}
\newcommand{\yk}{\kappa}
\newcommand{\defin}{\overset{\text{\tiny def}}{=}}
\newcommand{\ye}{\dens{\yesc}}
\newcommand{\yei}{\Tilde{\ye}}
\newcommand{\yeio}{\vac{\yei}}
\newcommand{\yesc}{E}
\newcommand{\yj}{\dens{\yjsc}}
\newcommand{\yji}{\Tilde{\yj}}
\newcommand{\yjio}{{\vac{\yji}}}
\newcommand{\yjsc}{J}
\newcommand{\yt}{X}
\newcommand{\ytp}{\Tilde{\yt}}
\newcommand{\ytf}{t}
\newcommand{\yas}{a}
\newcommand{\yden}{Q}

\newcommand{\ints}{\int_\ys}
\newcommand{\intp}{\int_\yp}

\DeclareMathOperator{\arcsinh}{arcsinh}
\DeclareMathOperator{\arccosh}{arccosh}

\DeclareMathOperator{\sgn}{sgn}
\newcommand{\tr}{\oper{tr}}
\newcommand{\lie}{\oper{L}}
\newcommand{\grad}{\oper{d}}

\begin{document}
\maketitle

\begin{abstract}
A generalization of two recently proposed general relativity
Hamiltonians, to the case of a general $(\yd\!+\!1)$-dimensional
dilaton gravity theory in a manifold with a timelike or spacelike
outer boundary, is presented.
\end{abstract}

\section{Introduction}\label{sec:introduction}

The study of Hamiltonians for general relativity and other gravity
theories is important for many interrelated questions and issues,
such as black hole thermodynamics, in particular black hole entropy
and its statistical origin, or as the definition of quasilocal
quantities. In particular, the boundary terms, which are part of the
Hamiltonian, are especially relevant. In fact, the Hamiltonian
reduces to them when evaluated on-shell and they are used to
determine global charges and thermodynamic quantities.

The form of the Hamiltonian boundary terms depends on the boundary
conditions we use for the variational principle (for instance we can
choose to fix the metric induced on the boundary), or on gauge conditions
such as, for instance, the orthogonality of the boundaries. In the
general framework of the Arnowitt-Deser-Misner parametrization, three
different gravitational Hamiltonians have been proposed recently by,
respectively, Hawking and Hunter (HH)~\cite{HH}, Booth and Mann
(BM)~\cite{BM1}, Creighton and Mann (CM)~\cite{CM}.

The HH and BM Hamiltonians  correspond
to different Legendre transformations of the 
Einstein-Hilbert action and represent the natural choice 
for classical general relativity defined  on a spacetime manifold
with non-orthogonal boundaries.
Conversely, the CM proposal gives  the  Hamiltonian for a 
dilaton gravity theory defined on a spacetime manifold with orthogonal
boundaries.

The limitations of the different proposal are evident. If one wants
to describe dilaton gravity theories in the Hamiltonian framework,
one has to use the CM prescription and is therefore forced to
consider only orthogonal boundaries. In some situation this
limitation may be too strong, for instance if one wants to consider
symmetry transformations whose generators cannot be tangent to a
timelike boundary (e.g.~spatial or null translations). This kind of
generators are important in the discussion of the asymptotic
symmetries of the spacetime and the associated charges~\cite{RT, BH,
CaMi, CaMi2}. On the other hand, if one needs to consider a manifold
with non-orthogonal boundaries, one can use the HH or BM prescription
but is limited to the non-dilatonic case.

Therefore, it is natural to investigate the possibility of using a
Booth-Mann-like Hamiltonian together with \emph{any} evolution
generator. In this paper we show that this is possible. We propose
two Hamiltonians for a general dilaton gravity theory defined on a
$(\yd\!+\!1)$-dimensional spacetime with non-orthogonal boundaries.
Our Hamiltonians generalize and comprehend the HH, BM, and CM
Hamiltonians. Moreover, they can deal with spacelike, as well as
timelike, outer boundaries.

The structure of the paper is the following. In Sect.~2 we set up our
notation and define the objects we are dealing with. In Sect.~3 we
derive our Hamiltonians. In Sect.~4 we discuss our results.

\section{Definitions}

We consider a $(\yd\!+\!1)$-dimensional spacetime manifold~$\ym$
whose boundary consists of two spacelike hypersurfaces~$\ys'$
and~$\ys''$ sharing the same topology, and an `outer'
hypersurface~$\yb$, which can be either timelike or spacelike, with
topology $\yf\ys'\times\yi$, where $\yi$ is a real interval. The
spacetime is foliated into spacelike hypersurfaces~$\ys_t$ of
constant~$t$, where $\ytf:\ym\rightarrow\R$~is a time function
defined throughout~$\ym$. The initial and final hypersurfaces of this
foliation are~$\ys'$ and~$\ys''$. Another foliation is induced on the
boundary~$\yb$ and is given by spacelike surfaces~$\yp_t = \ys_t \cap
\yb=\yf\ys_t$ of dimension~$(\yd\!-\!1)$. The initial and final
surfaces are~$\yp' = \ys' \cap \yb$ and~$\yp'' = \ys'' \cap \yb$,
respectively. We can also think of every~$\yp_t$ as given by the
intersection of~$\yb$ with (local) orthogonal
hypersurfaces~$\ysp_t$.%
\footnote{When the~$\yb$ boundary is spacelike, we only assume
locality and do not suppose that the hypersurfaces~$\ysp_t$ foliate
the whole spacetime~$\ym$. In this case $\ysp_t$ is timelike and
we do not want a foliation of~$\ym$ into timelike hypersurfaces.
Thus, \emph{clamped foliations}, in the sense of
Lau~\cite{Lau1,Lau2}, are allowed only when~$\yb$ is timelike.}

When the boundary~$\yb$ is spacelike, it must also be transversal to
every~$\ys_t$ to be sure that the foliation of~$\ym$ is always
well-defined. However, later on we will relax this assumption and
will allow for the degenerate case in which~$\yb$ is non-transversal
to some or every~$\ys_t$.

The metric on the spacetime~$\ym$ is~$g_{\yim\yjm}$, with
signature~$(-++\cdots)$, volume element~$\ygmd$, covariant
derivative~$\ydm_\yim$ and curvature~$\yrm$. With respect to this
metric the $\ym$-foliation lapse
is~$\yns\defin[-(\ydm\ytf)^2]^{-1/2}$. The metric $g_{\yim\yjm}$
induces other metric structures on the various surfaces. These
structures are described in detail below.

A scalar dilaton field~$\dil$ is also defined on~$\ym$, as well as
its functions~$\fdil:\dil\mapsto\fdil(\dil)$,
$\kdil:\dil\mapsto\kdil(\dil)$
and~$\pdil:\dil\mapsto\pdil(\dil)$; their
derivatives~$\frac{d\fdil}{d\dil}$ etc.\ are written as~$\fdil'$
etc.;\
their restrictions to the various surfaces, $\dil|_{\ys_t}$,
$\dil|_{\yb}$, $\fdil|_{\yp_t}$, etc.,\ will be often called~$\dil$,
$\fdil$, etc.\ for simplicity.

The Lie derivative operator is denoted by~$\lie$.


\subsubsection{The~$\ys_t$ hypersurfaces}

A future-pointing vector field~$\yus^\yim$ normal to every~$\ys_t$ is
defined on~$\ym$; its acceleration~$\yas^\yim \defin \yus^\yjm
\ydm_\yjm \yus^\yim$ is tangent to~$\ys_t$~($\yus_\yim \yas^\yim=0$).
The induced Riemannian metric on every~$\ys_t$
is~$\ygs_{\yim\yjm}=g_{\yim\yjm}+\yus_\yim \yus_\yjm$, with volume
element~$\ygsd$ (note that~$\ygsd\yns=\ygmd$), covariant
derivative~$\yds_\yis$, intrinsic curvature~$\yrs$, and extrinsic
curvature~$\ycs_{\yim\yjm}\defin -{\ygs_\yim}^\ykm \ydm_\ykm
\yus_\yjm$. Tensors are projected onto the~$\ys_t$ hypersurfaces
by~${\ygs^\yim}_\yjm$.

\subsubsection{The~$\yb$ boundary}

The outer boundary~$\yb$, whose normal~$\yub^\yim$ we require to be
always outward-pointing, can be timelike (see Fig.~\ref{fig:M1}) or
spacelike. When $\yb$ is spacelike we have to distinguish between two
different cases, sketched in Figs.~\ref{fig:M2} and~\ref{fig:M3}. In
Fig.~\ref{fig:M2} $\yb$~lies outside the future of~$\ys'$ (its normal
is past-pointing), while it lies inside the future of~$\ys'$ in
Fig.~\ref{fig:M3} (its normal is future-pointing). We can
characterize the three different cases defining the following
quantities:
\begin{align}
\ep&\defin\yub_\yim\yub^\yim,\label{eq:epsilon}\\
\ypz&\defin\yus^{\yim} \yub_{\yim},\label{eq:beta}\\
\intertext{as well as the hyperbolic angle~$\ypa$:}
\ypa&\defin
\begin{cases}
\arcsinh\ypz & \text{if $\ep=+1$},\\
-\sgn(\ypz) \arccosh|\ypz| & \text{if $\ep=-1$}.
\end{cases}\label{eq:angle}
\end{align}

If~$\yb$ is timelike (see Fig.~\ref{fig:M1})
then~$\ep=+1$,~$\ypz\gtreqless 0$, and~$\ypa\gtreqless 0$; if~$\yb$
is spacelike and outside the future of~$\ys'$
(past-pointing~$\yub^\yim$, see Fig.~\ref{fig:M2})
then~$\ep=-1$,~$\ypz> 1$, and~$\ypa < 0$; finally, if~$\yb$ is
spacelike and inside the future of~$\ys'$
(future-pointing~$\yub^\yim$, see Fig.~\ref{fig:M3}) then~$\ep =
-1$,~$\ypz < -1$, and~$\ypa > 0$.

\begin{figure}[!p]
\begin{center}
\setlength{\unitlength}{1pt}
\begin{picture}(260,80)(-55,-40)
\thinlines
\put(20,15){\makebox(0,0){$\ym$}}
\put(20,-30){\line(1,0){110}}\put(75,-31){\makebox(0,0)[t]{$\ys'$}}
\put(0,0){\line(1,0){150}}\put(75,1){\makebox(0,0)[b]{$\ys_t$}}
\put(-20,30){\line(1,0){190}}\put(75,31){\makebox(0,0)[b]{$\ys''$}}
\multiput(20,-30)(-3,3){21}{\circle*{.1}}
\multiput(20,-30)(3,3){6}{\circle*{.1}}
\multiput(130,-30)(3,3){21}{\circle*{.1}}
\multiput(130,-30)(-3,3){6}{\circle*{.1}}
\multiput(150,0)(2,2){10}{\circle*{.1}}
\multiput(150,0)(-2,2){10}{\circle*{.1}}
\multiput(150,0)(2,-2){10}{\circle*{.1}}
\multiput(150,0)(-2,-2){10}{\circle*{.1}}
\put(20,-30){\line(-2,3){40}}
\put(130,-30){\line(2,3){40}}\put(130,-10){\makebox(0,0)[tl]{$\yb$}}
\put(20,-30){\circle*{2}}\put(130,-30){\circle*{2}}
\put(130,-31){\makebox(0,0)[t]{$\yp'$}}
\put(0,0){\circle*{2}}\put(150,0){\circle*{2}}
\put(150,-1){\makebox(0,0)[tl]{$\yp_t$}}
\put(-20,30){\circle*{2}}\put(170,30){\circle*{2}}
\put(170,31){\makebox(0,0)[b]{$\yp''$}}
\thicklines
\put(150,0){\vector(1,0){20}}\put(170,0){\makebox(0,0)[l]{$\yups^\mu$}}
\put(150,0){\vector(0,1){20}}\put(150,20){\makebox(0,0)[r]{$\yus^\mu$}}
\put(150,0){\vector(2,3){17.89}}\put(170,30){\makebox(0,0)[lt]{$\yupb^\mu$}}
\put(150,0){\vector(3,2){26.83}}\put(180,20){\makebox(0,0)[lt]{$\yub^\mu$}}
\end{picture}
\caption{\small Example of foliation of a two-dimensional
spacetime~$\ym$ with a timelike outer boundary
($\ep\!=\!+1$,~$\ypz\!\gtreqless\! 0$,~$\ypa\!\gtreqless\! 0$). Dotted lines
represent lightcones.}
\label{fig:M1}
\end{center}
\end{figure}
\begin{figure}[!p]
\begin{center}
\setlength{\unitlength}{1pt}
\begin{picture}(260,80)(-55,-40)
\thinlines
\put(45,15){\makebox(0,0){$\ym$}}
\put(45,-30){\line(1,0){60}}\put(75,-31){\makebox(0,0)[t]{$\ys'$}}
\put(0,0){\line(1,0){150}}\put(75,1){\makebox(0,0)[b]{$\ys_t$}}
\put(-45,30){\line(1,0){240}}\put(75,31){\makebox(0,0)[b]{$\ys''$}}
\multiput(45,-30)(-3,3){21}{\circle*{.1}}
\multiput(45,-30)(3,3){6}{\circle*{.1}}
\multiput(105,-30)(3,3){21}{\circle*{.1}}
\multiput(105,-30)(-3,3){6}{\circle*{.1}}
\multiput(150,0)(2,2){10}{\circle*{.1}}
\multiput(150,0)(-2,2){10}{\circle*{.1}}
\multiput(150,0)(2,-2){10}{\circle*{.1}}
\multiput(150,0)(-2,-2){10}{\circle*{.1}}
\put(45,-30){\line(-3,2){90}}
\put(105,-30){\line(3,2){90}}\put(120,-20){\makebox(0,0)[br]{$\yb$}}
\put(45,-30){\circle*{2}}\put(105,-30){\circle*{2}}
\put(105,-31){\makebox(0,0)[t]{$\yp'$}}
\put(0,0){\circle*{2}}\put(150,0){\circle*{2}}
\put(150,-1){\makebox(0,0)[tl]{$\yp_t$}}
\put(-45,30){\circle*{2}}\put(195,30){\circle*{2}}
\put(195,31){\makebox(0,0)[b]{$\yp''$}}
\thicklines
\put(150,0){\vector(1,0){20}}\put(170,0){\makebox(0,0)[l]{$\yups^\mu$}}
\put(150,0){\vector(0,1){20}}\put(150,20){\makebox(0,0)[b]{$\yus^\mu$}}
\put(150,0){\vector(-2,-3){17.89}}\put(130,-30){\makebox(0,0)[tl]{$\yub^\mu$}}
\put(150,0){\vector(3,2){26.83}}\put(180,20){\makebox(0,0)[tl]{$\yupb^\mu$}}
\end{picture}
\caption{\small Example of foliation of a two-dimensional
spacetime~$\ym$ with a spacelike outer boundary outside the future
of~$\ys'$ ($\ep\!=\!-1$,~$\ypz\!>\! +1$,~$\ypa\!<\! 0$). Dotted lines
represent lightcones.}
\label{fig:M2}
\end{center}
\end{figure}

\begin{figure}[!p]
\begin{center}
\setlength{\unitlength}{1pt}
\begin{picture}(260,80)(-55,-40)
\thinlines
\put(45,15){\makebox(0,0){$\ym$}}
\put(45,30){\line(1,0){60}}\put(75,31){\makebox(0,0)[b]{$\ys''$}}
\put(0,0){\line(1,0){150}}\put(75,1){\makebox(0,0)[b]{$\ys_t$}}
\put(-45,-30){\line(1,0){240}}\put(75,-31){\makebox(0,0)[t]{$\ys'$}}
\multiput(-45,-30)(-3,3){6}{\circle*{.1}}
\multiput(-45,-30)(3,3){21}{\circle*{.1}}
\multiput(195,-30)(3,3){6}{\circle*{.1}}
\multiput(195,-30)(-3,3){21}{\circle*{.1}}
\multiput(150,0)(2,2){10}{\circle*{.1}}
\multiput(150,0)(-2,2){10}{\circle*{.1}}
\multiput(150,0)(2,-2){10}{\circle*{.1}}
\multiput(150,0)(-2,-2){10}{\circle*{.1}}
\put(45,30){\line(-3,-2){90}}
\put(105,30){\line(3,-2){90}}\put(164,-10){\makebox(0,0)[tr]{$\yb$}}
\put(45,30){\circle*{2}}\put(105,30){\circle*{2}}
\put(105,31){\makebox(0,0)[b]{$\yp''$}}
\put(0,0){\circle*{2}}\put(150,0){\circle*{2}}
\put(150,-1){\makebox(0,0)[tr]{$\yp_t$}}
\put(-45,-30){\circle*{2}}\put(195,-30){\circle*{2}}
\put(195,-31){\makebox(0,0)[t]{$\yp'$}}
\thicklines
\put(150,0){\vector(1,0){20}}\put(170,0){\makebox(0,0)[lb]{$\yups^\mu$}}
\put(150,0){\vector(0,1){20}}\put(150,20){\makebox(0,0)[bl]{$\yus^\mu$}}
\put(150,0){\vector(-2,3){17.89}}\put(130,30){\makebox(0,0)[b]{$\yub^\mu$}}
\put(150,0){\vector(-3,2){26.83}}\put(120,20){\makebox(0,0)[tr]{$\yupb^\mu$}}
\end{picture}
\caption{\small Example of foliation of a two-dimensional
spacetime~$\ym$ with a spacelike outer boundary inside the future
of~$\ys'$ ($\ep\!=\!-1$,~$\ypz\!<\! -1$,~$\ypa\!>\! 0$). Dotted
lines represent lightcones.}
\label{fig:M3}
\end{center}
\end{figure}

On~$\yb$ we have the induced intrinsic metric~$\ygb_{\yim\yjm}=
g_{\yim\yjm} -\ep \yub_\yim \yub_\yjm$ with volume element~$\ygbd$,
covariant derivative~$\ydb_\yib$ and extrinsic
curvature~$\ycb_{\yib\yjb}\defin - {\ygb_\yib}^\ykm \ydm_\ykm
\yub_\yjb$. In the $\ep=+1$~case the induced metric is Lorentzian
with signature~$(-++\cdots)$, while in the $\ep=-1$~case it is a
positive definite Riemannian metric having signature~$(+++\cdots)$.
We can project tensors onto~$\yb$ by using~${\ygb^\yim}_\yjm$.

The spacetime foliation induces a foliation in~$\yb$ by means of the
induced time function~$\ytf|_\yb:\yb\rightarrow\R$, and the
associated lapse is~$\ynp\defin[-\ep(\ydb\ytf|_\yb)^2]^{-1/2}$.

\subsubsection{The~$\yp_t$ surfaces}

The~$\yp_t$ surfaces are defined by the intersection of the outer
boundary with the various slices~$\ys_t$, so they can be viewed as
embedded in~$\yb$ or in~$\ys_t$. In particular,~$\yp'$ and~$\yp''$
together form the boundary of~$\yb$, and every~$\yp_t$ is the
boundary of~$\ys_t$. Hence, four different unit normal vector fields
can be defined on~$\yp_t$: as a surface in~$\ys_t$, $\yp_t$~has the
outward-pointing spacelike normal~$\yups^\yim$, and shares
with~$\ys_t$ the future-pointing timelike unit normal~$\yus^\yim$; as
a surface in~$\yb$, $\yp_t$~has the unit normal~$\yupb^\yim$, which
satisfies~$\yus_\yim\yupb^\yim < 0$, and shares with~$\yb$
the outward-pointing unit normal~$\yub^\yim$. Both~$\yups^\yim$
and~$\yupb$ can be obtained by projection of the normals~$\yub^\yim$
and~$\yus^\yim$ onto~$\ys_t$ and onto~$\yb$ respectively, and
normalizing,
\begin{subequations}\label{eq:norm0}
\begin{align}
\yups^\yim&= \ep \ypl {\ygs^\yim}_\yjm \yub^\yim
=\ep\ypl\yub^\yim +\ep\ypl\ypz\yus^\yim, \\ 
\yupb^\yim&= \ep \ypl {\ygb^\yim}_\yjm\yus^\yim
=\frac{1}{\ypl}\yus^\yim -\ypl\ypz\yub^\yim,
\end{align}
\end{subequations}
where the normalizing positive scalar~$\ypl$ is defined by:
\begin{equation}\label{eq:lambda}
\ypl=\frac{1}{\sqrt{\ep +\ypz^2}}
=\begin{cases}
(\cosh\ypa)^{-1} &   \text{if $\ep=+1$},\\
(\sinh|\ypa|)^{-1}  &  \text{if $\ep=-1$}.
\end{cases}
\end{equation}
Note that~$\var\ypz=\ep\frac{1}{\ypl} \var\ypa$;
moreover,~$\yups_\yim\yups^\yim\!=\!+1$, $\yus_\yim\yus^\yim\!=\!-1$,
$\yupb_\yim\yupb^\yim\!=\!-\ep$, $\yub_\yim\yub^\yim\!=\!\ep$, and
the following relations hold:
\begin{subequations}\label{eq:norm1}
\begin{align}
\yub^\yim&=\ep\frac{1}{\ypl}\yups^\yim-\ypz\yus^\yim&
\yupb^\yim&=\frac{1}{\ypl}\yus^\yim -\ep\ypz\yups^\yim,\label{eq:norm1tilde}\\
\yups^\yim&=\frac{1}{\ypl}\yub^\yim +\ypz\yupb^\yim&
\yus^\yim&=\ep\frac{1}{\ypl}\yupb^\yim +\ep\ypz\yub^\yim.\label{eq:norm1nont}
\end{align}
\end{subequations}
The Riemannian metric induced on~$\yp_t$ is
\begin{equation}\label{eq:metrP}
\ygp_{\yim\yjm} =g_{\yim\yjm} -\yups_\yim \yups_\yjm + \yus_\yim \yus_\yjm =
g_{\yim\yjm} +\ep\yupb_\yim \yupb_\yjm -\ep \yub_\yim \yub_\yjm,
\end{equation}
with volume element~$\ygpd$ (and~$\ygpd\ynp=\ygbd$). Tensors are
projected on~$\yp_t$ by using~${\ygp^\yim}_\yjm$.

Every~$\yp_t$ can be seen as a $(\yd\!-\!1)$-dimensional surface
embedded in~$\ys_t$, and as such it has an extrinsic curvature which
we denote by~$\ycps_{\yip\yjp}$:
\begin{equation}\label{eq:curvPS}
\ycps_{\yip\yjp}\defin -\ygp_\yip^\ykm \ygp_\yjp^\ylm
\ydm_\ykm\yups_\ylm = -\ygp_\yip^\yks \yds_\yks\yups_\yjp.
\end{equation}

However, we can consider also the embedding of~$\yp_t$ in a
hypersurface~$\ysp_t$, locally orthogonal to~$\yb$ (so that~$\ysp_t$
has unit normal~$\yupb^\yim$); in this case we can define an associated 
extrinsic curvature~$\ycpb_{\yip\yjp}$:
\begin{equation}\label{eq:curvPB}
\ycpb_{\yip\yjp}\defin -\ygp_\yip^\ykm \ygp_\yjp^\ylm
\ydm_\ykm\yub_\ylm.
\end{equation}

The following useful relation holds among the traces
of~$\ycps_{\yip\yjp}$,~$\ycpb_{\yip\yjp}$ and the extrinsic
curvatures of~$\ys_t$ and~$\yb$:
\begin{align}
\tr\ycps &= \ep \ypl \tr\ycb + \ep\ypl\ypz\tr\ycs + \ep\ypl\yub^\yim \yas_\yim
-\ypl\yupb^\yim\ydm_\yim \ypa,
\label{eq:trps1}\\
& =\ep \tr\ycpb + \ep\ypl\ypz\tr\ycs + \ep\ypl\ypz\yups^\yim \yub^\yjm
\ydm_\yim \yupb_\yjm + \ep \ypl\ypz \yups^\yim \ydm_\yim \ypa.
\label{eq:trps2}
\end{align}

\subsubsection{Bulk and boundary foliations}

The time evolution of the hypersurfaces~$\ys_t$ (and of the fields
defined on them) can be specified by means of a time-flow vector
field~$\yt^\yim$, satisfying~$\grad\ytf(\yt)\equiv1$. An equivalent 
definition is:
\begin{equation}\label{eq:time1}
\yt^\yim =\yns\yus^\yim +\yvs^\yim, 
\end{equation}
where~$\yns\equiv[-(\ydm\ytf)^2]^{-1/2}=-\yus_\yim \yt^\yim$
and~$\yvs^\yis\equiv{\ygs^\yis}_\yjm \yt^\yjm$ are the (bulk) lapse
and shift, respectively.

Analogously, the time evolution of the surfaces~$\yp_t$ along~$\yb$
can be specified by a boundary time-flow vector
field~$\ytp^\yib$ tangent to~$\yb$ and such that~$\grad\ytf|_\yb
(\ytp)\equiv1$. We have now,
\begin{equation}\label{eq:time1p}
\ytp^\yib =\ynp\yupb^\yib +\yvp^\yib,
\end{equation}
where~$\ynp\equiv[-\ep(\ydb\ytf|_\yb)^2]^{-1/2}=-\ep\yupb_\yib
\ytp^\yib$ and~$\yvp^\yip\equiv{\ygb^\yip}_\yjb \ytp^\yjb$ are the
\emph{boundary} lapse and shift.

In general, the bulk time-flow vector field~$\yt^\yim$ is not tangent
to the outer boundary,~$\yub_\yim \yt^\yim \neq 0$, so that it differs
from the boundary time-flow vector field:~$\yt^\yim|_\yb\neq\ytp^\yim$.
This means that the bulk and boundary shifts are unrelated to each
other. Note, though, that the bulk and boundary lapses~$\yns$
and~$\ynp$ are always related by:
\begin{equation}\label{eq:shift1}
\ynp =\ypl \yns.
\end{equation}
This equation is just a consequence of the fact that the $\yb$~foliation is
induced by the $\ym$~foliation.

When the vector field~$\yt^\yim$ is tangent
to~$\yb$,~$\yub_\yim \yt^\yim = 0$, we may require the two time-flow
vector fields to coincide,~$\yt^\yim|_\yb=\ytp^\yim$, so that the
respective shifts are related by
\begin{equation}\label{eq:shift2}
\yvp^\yim = {\ygp^\yim}_\yjm \yvs^\yjm
= \yvs^\yim + \ep\ynp\ypz \yups^\yim.
\end{equation}

\subsubsection{The general action}

The action for a general dilaton gravity theory on a
$(\yd\!+\!1)$-dimensional spacetime is
\begin{equation}
  \label{eq:lag}
\begin{split}
  \yli \defin &\frac{1}{2\yk} \int_\ym \ygmd [\fdil(\dil) \yrm +
  \kdil(\dil) (\ydm\dil)^2 + \pdil(\dil)] \\
& +  \frac{1}{\yk} \ints \ygsd \fdil(\dil) \tr\ycs 
- \frac{\ep}{\yk}\int_{\yb} \ygbd \fdil(\dil) \tr\ycb
+ \frac{1}{\yk} \intp \ygpd \fdil(\dil) \ypa,
\end{split}
\end{equation}
where~$\int_\ys$ and~$\int_\yp$ are abbreviations
for~$\int_{\ys''}-\int_{\ys'}$ and~$\int_{\yp''}-\int_{\yp'}$
respectively. The boundary terms make the action suitable for a
variational principle with Dirichlet boundary conditions, i.e.~with
boundary-induced metric and dilaton field fixed.

The terms~$\fdil$, $\kdil$, $\pdil$, and the constant~$\yk$ depend on
the model under consideration (for example, setting~$\yd = 1$, $\yk =
\pi$, $\dil=e^{-2\phi}$, $\fdil(\dil)=\dil$, $\kdil(\dil)=0$,
$\pdil(\dil)=-2\Lambda\dil$, we have the Jackiw-Teitelboim
action~\cite{JT,BL}; setting~$\yd=3$, $\yk=8\pi G$,
$\dil=\fdil\equiv1$, $\kdil=\pdil\equiv0$, we have the classical
Einstein-Hilbert action~\cite{BY,BLY}; cf.~also Creighton
and~Mann~\cite{CM} and~Lemos~\cite{Le}). We have not included
minimally-coupled matter terms in the action, because the presence of
these terms does not affect our main results.

The variation of the action~(\ref{eq:lag}) is:
\begin{equation}
\label{eq:lagvar}
\begin{split}
\var\yli = & \int_\ym (\ypm^{\yim\yjm} \var g_{\yim\yjm} +
\ypmdil \var\dil ) 
+ \ints (\yps^{\yis\yjs} \var\ygs_{\yis\yjs} +
\ypsdil \var\dil ) \\
&+ \int_\yb (\ypb^{\yib\yjb} \var\ygb_{\yib\yjb} +
\ypbdil \var\dil ) 
+ \intp (\ypp^{\yip\yjp} \var\ygp_{\yip\yjp} +
\yppdil \var\dil ),
\end{split}
\end{equation}
where
\begin{subequations}\label{eq:momm}
\begin{align}
\begin{split}
\ypm^{\yim\yjm} =& 
-\frac{1}{2\yk} \ygmd [
\fdil \yrm^{\yim\yjm} -
\tfrac{1}{2} \yrm g^{\yim\yjm} - \tfrac{1}{2} \pdil g^{\yim\yjm}
+g^{\yim\yjm} \ydm^2\fdil
\\
&\qquad- \ydm^{\yim} \ydm^{\yjm}\fdil + \kdil
\ydm^{\yim}\dil \ydm^{\yjm}\dil - \tfrac{1}{2} \kdil g^{\yim\yjm}
(\ydm\dil)^2],
\end{split}
\\
\ypmdil = & \frac{1}{2\yk} \ygmd [ \fdil' \yrm +\pdil' + \kdil'
(\ydm^2 \dil) - 2 \ydm^{\yim} (\kdil
\ydm_{\yim}\dil) ]
\end{align}
\end{subequations}
are the terms which give the equations of motion of the
theory,~$\ypm^{\yim\yjm}=0$ and~$\ypmdil=0$; whereas
\begin{subequations}\label{eq:moms}
\begin{align}
\yps^{\yim\yjm}= &- \frac{\ygsd}{2\yk} [ \fdil (\ycs^{\yim\yjm} -
\tr\ycs \ygs^{\yim\yjm}) + \ygs^{\yim\yjm} \lie_\yus \fdil],
\label{eq:moms-a}\\ 
\ypsdil= &\frac{\ygsd}{\yk} (\fdil' \tr\ycs - \kdil
\lie_\yus\dil)
\end{align}
\end{subequations}
are the momenta conjugated to~$\ygs_{\yis\yjs}$ and~$\dil|_\ys$ on~$\ys$;
\begin{subequations}\label{eq:momb}
\begin{align}
\ypb^{\yim\yjm}&= \frac{\ep}{2\yk}\ygbd [ \fdil
(\ycb^{\yim\yjm} - \tr\ycb \ygb^{\yim\yjm}) + \ygb^{\yim\yjm}
\lie_\yub \fdil],\\
\ypbdil&= -\frac{\ep}{\yk}\ygbd (\fdil' \tr\ycb
- \kdil \lie_\yub\dil)
\end{align}
\end{subequations}
are the momenta conjugate to~$\ygb_{\yib\yjb}$ and~$\dil|_\yb$
on~$\yb$; and finally
\begin{subequations}\label{eq:momp}
\begin{align}
\ypp^{\yip\yjp}&=\frac{1}{2\yk}\ygpd\fdil\ypa\ygp^{\yip\yjp},\\
\yppdil&=\frac{1}{\yk}\ygpd\fdil'\ypa
\end{align}
\end{subequations}
are the momenta conjugate to~$\ygp_{\yip\yjp}$ and~$\dil|_\yp$
on~$\yp$.

\section{Derivation of the Hamiltonians}

In this section we derive two Hamiltonians, corresponding to two
different Legendre transformations of the action~(\ref{eq:lag}),
which are the generalizations for a dilaton gravity theory of those
proposed by Hawking and Hunter~\cite{HH}, and Booth and
Mann~\cite{BM1}.

\subsubsection{First Hamiltonian}
The action~(\ref{eq:lag}) is first decomposed with respect to the
foliation in the standard way, using the Gauss-Codazzi equations
\begin{align}
\yrm &= \yrs + \ycs^{\yis\yjs}
\ycs_{\yis\yjs} - (\tr\ycs)^2 - 2 \ydm_{\yim}(\yus^{\yim} \tr\ycs +
\yas^{\yim}),\\
\intertext{and the decomposition of the squared divergence of
the dilaton}
(\ydm \dil)^2 &= (\yds\dil)^2 - (\lie_\yus \dil)^2.
\end{align}
One  obtains:
\begin{equation}
\begin{split}
\yli = & \frac{1}{2\yk} \int_\ym \ygmd \{
\fdil\yrs +
\fdil \ycs^{\yis\yjs} \ycs_{\yis\yjs} - \fdil(\tr\ycs)^2
+\kdil(\yds\dil)^2 - \kdil(\lie_\yus\dil)^2 \\
&\qquad+\pdil +2 \tr\ycs \lie_\yus\fdil +2 \lie_\yas\fdil
- 2\ydm_{\yim}[\fdil(\yus^{\yim} \tr\ycs +
\yas^{\yim})]\}
\\
& +  \frac{1}{\yk} \ints \ygsd \fdil \tr\ycs 
- \frac{\ep}{\yk}  \int_{\yb} \ygbd \fdil \tr\ycb
+ \frac{1}{\yk} \intp \ygpd \fdil \ypa.
\end{split}
\end{equation}

We can rewrite the intrinsic curvature~$\ycs^{\yim\yjm}$ and the Lie
derivative of the dilaton~$\lie_\yus\dil$ in terms of the
momenta~$\yps^{\yim\yjm}$ and~$\ypsdil$:
\begin{subequations}\label{eq:tomoms}
\begin{align}
\ycs^{\yim\yjm}=& \frac{\yk}{\fdil\yden\ygsd}\{
-2\yden \yps^{\yim\yjm} 
+2 [(\fdil')^2 - \fdil\kdil]\tr\yps \ygs^{\yim\yjm}
+\fdil\fdil' \ypsdil\ygs^{\yim\yjm}\},\\
\lie_\yus\dil =& \frac{\yk}{\yden\ygsd}
[-2\fdil' \tr\yps + (\yd-1)\fdil \ypsdil],
\end{align}
\end{subequations}
with
\begin{equation}\label{eq:den}
\yden\defin \yd(\fdil')^2 - (\yd-1)\fdil\kdil.
\end{equation}
Using the previous  equations, together with the relation
\begin{equation}
\lie_\yas\fdil =\frac{1}{\yns} \yds_\yis (\yns
\yds^\yis \fdil) - \yds^2\fdil,
\end{equation}
we get
\begin{equation}\label{eq:lag-fol}
\begin{split}
\yli = &
\begin{aligned}[t]
 \int_{\ym} \biggl[&
\yps^{\yis\yjs} \lie_{\yt}\ygs_{\yis\yjs} + 
\ypsdil \lie_{\yt}\dil - \yns\yhd - \yvs^\yis \ymd_\yis\\
&- \frac{1}{\yk}\ygmd\ydm_{\yim}(\yus^{\yim} \fdil\tr\ycs 
+ \yas^{\yim}\fdil)\\
&+\frac{1}{\yk}\ygsd\yds_\yis \biggl(\yns \yds^\yis \fdil
-2\yvs_\yjs \frac{\yk}{\ygsd}\yps^{\yis\yjs}\biggr)\biggr]
\end{aligned}\\
& +  \frac{1}{\yk} \ints \ygsd \fdil \tr\ycs 
- \frac{\ep}{\yk}  \int_{\yb} \ygbd \fdil \tr\ycb
+ \frac{1}{\yk} \intp \ygpd \fdil \ypa,
\end{split}
\end{equation}
where the Hamiltonian constraints~$\yhd$ and~$\ymd_\yis$ are given by
\begin{subequations}
\begin{align}
\begin{split}
\yhd\defin&
\frac{2\yk}{\ygsd}\biggl[\frac{1}{\fdil}\yps^{\yis\yjs}\yps_{\yis\yjs}
- \frac{(\fdil')^2 - \fdil\kdil}{\fdil \yden} (\tr\yps)^2
- \frac{\fdil'}{\yden} \tr\yps \ypsdil  \\
&\quad+ \frac{(\yd - 1)\fdil}{4\yden} (\ypsdil)^2\biggr] 
-\frac{\ygsd}{2\yk}[\fdil\yrs +\kdil(\yds\dil)^2 +\pdil -2\yds^2\fdil],
\end{split}\\
\ymd_\yis\defin& -2\yds_\yjs {\yps^\yjs}_\yis + \ypsdil \yds_\yis\dil,
\end{align}
\end{subequations}
and~$\yden$ is defined in Eq.~(\ref{eq:den}).

Let us now focus on the additional boundary terms in
Eq.~(\ref{eq:lag-fol}) that represent total derivatives. The first
term yields boundary terms on~$\ys'$, $\ys''$, and~$\yb$:
\begin{equation}
\begin{split}
\yli_{1}&=-\frac{1}{\yk}\int_{\ym} \ygmd\ydm_{\yim}(\yus^{\yim}
\fdil\tr\ycs +\yas^{\yim}\fdil)\\
&= \frac{1}{\yk} \ints \ygsd (-\fdil
\tr\ycs + \yus_\yis \yas^\yis\fdil)
 - \frac{\ep}{\yk}\int_\yb \ygbd (\fdil\ypz \tr\ycs
 + \fdil \yub^\yim \yas_\yim).
\end{split}
\end{equation}
Since~$\yas^\yis$ is orthogonal to~$\yus^\yis$, we see that the first
integral in~$\yli_{1}$ exactly
cancels out with the~$\ys$-integral already present in the
action~(see Eq.~(\ref{eq:lag})). The second integral, instead, sums up
with the~$\yb$-integral of the action to give
\begin{equation}
\label{e1}
\yli_{2}=-\frac{\ep}{\yk}\int_{\yb}\ygbd(\fdil\tr\ycb
+\fdil\ypz\tr\ycs + \fdil \yub_\yim  \yas^\yim).
\end{equation}
Using Eq.~(\ref{eq:trps1}) one can show that Eq.~(\ref{e1}) can be 
written as

\begin{equation}
\yli_{2}=-\frac{1}{\yk}\int_{\yb}\ygbd(\fdil\ypl^{-1}\tr\ycps 
+ \fdil \yupb^\yib \ydb_\yib\ypa).
\end{equation}
Taking out of it a total divergence by using~$\fdil \yupb^\yib
\ydb_\yib\ypa = \ydb_\yib (\fdil \yupb^\yib\ypa) - \ypa
\ydb_\yib(\fdil\yupb^\yib)$, one finds
\begin{equation}
\yli_{2}=-\frac{1}{\yk}\int_{\yb}\ygbd[\fdil\tr\ycps -\ypa \ydb_\yib(\fdil
\yupb^\yib)] -\frac{1}{\yk}\intp\ygpd\fdil\ypa,
\end{equation}
so that the $\yp$-integral in~$\yli_{2}$ exactly cancels out with the
$\yp$-integral which appears in the action (\ref{eq:lag}). Let us now
consider the last divergence in Eq.~(\ref{eq:lag-fol}). This term
yields the following boundary contribution:
\begin{equation}
\begin{split}
\yli_{3}&=\frac{1}{\yk}\int_t \int_{\ys_t}\ygsd \yds_\yis \biggl(\yns \yds^\yis \fdil
-2\yvs_\yjs \frac{\yk}{\ygsd}\yps^{\yis\yjs}\biggr)\\
&=\frac{1}{\yk}\int_t\int_{\yp_t}\ygsd \biggl(\yns \yups^\yis \yds_\yis\fdil
-2\yups_\yis \yvs_\yjs \frac{\yk}{\ygsd}\yps^{\yis\yjs}\biggr).
\end{split}
\end{equation}
It follows  that the action put into canonical form contains only
a boundary integral on~$\yb$:
\begin{equation}
\begin{split}
\yli = & \int_t\biggl\{\int_{\ys_t} ( \yps^{\yis\yjs}
\lie_{\yt}\ygs_{\yis\yjs} + \ypsdil \lie_{\yt}\dil - \yns\yhd -
\yvs^\yis \ymd_\yis)\\
 &-\frac{1}{\yk}\int_{\yp_t}\ygpd\biggl[\yns[\fdil\tr\ycps -
\ypl\ypa\ydb_\yib(\fdil \yupb^\yib) 
- \yups^\yim \yds_\yim\fdil]
- 2\yvs_\yis \yups_\yjs\frac{\yk}{\ygsd}\yps^{\yis\yjs}\biggr]\biggr\}.
\end{split}
\end{equation}

It is now straightforward to perform the Legendre transformation,
\begin{equation}\label{eq:Leg1}
\int_t\yhi \equiv \int_t\int_{\ys_t} 
(\yps^{\yis\yjs} \lie_{\yt}\ygs_{\yis\yjs} + 
\ypsdil \lie_{\yt}\dil) - \yli,
\end{equation}
which gives us the Hamiltonian:
\begin{equation}\label{eq:ham1}
\yhi= 
\int_{\ys_t} (\yns\yhd + \yvs^\yis \ymd_\yis)
+\int_{\yp_t} (\yns\ye - \yvs^\yis \yj_\yis),
\end{equation}
where
\begin{subequations}\label{eq:ejS}
\begin{align}
\ye\defin & \frac{\ygpd}{\yk}[\fdil\tr\ycps - \lie_\yups\fdil 
- \ypl\ypa\ydb_\yib(\fdil\yupb^\yib)], \\
\yj_\yis\defin& - 2\yups_\yjs \frac{\ygpd}{\ygsd}\yps^{\yis\yjs}.
\end{align}
\end{subequations}

Note that the term which depends on the hyperbolic angle~$\ypa$ can
be rewritten as follows by means of Eqs.~(\ref{eq:norm1tilde}),
(\ref{eq:metrP}), (\ref{eq:curvPS}), and~(\ref{eq:tomoms}):
\begin{equation}\label{eq:depangle}
\yns\ypl\ypa\ydb_\yib(\fdil\yupb^\yib)=
\yns\ypa\biggl[\ep\ypl\ypz(\fdil\tr\ycps - \lie_\yups\fdil)
-2\yk \yups_\yim\yups_\yjm \frac{\yps^{\yim\yjm}}{\ygsd}\biggr].
\end{equation}
The appearance of a term containing the intersection angle~$\ypa$,
which in this case does not depend on the canonical variables, is
analogous to what happens for the HH Hamiltonian. In fact,
the~$\yns\ye$ integral reduces, in the non-dilatonic case, to the sum
of the HH `curvature' and `tilting' terms, whereas the
$\yvs^\yis\yj_\yis$~integral reduces to the HH `momentum' term. In
order to get rid of the explicit angle dependence we have to subtract
a reference term from the Hamiltonian.

\subsubsection{Second Hamiltonian}

In this subsection we use Booth and Mann's prescription, i.e.~we
require the time-flow vector field~$\yt^\yim$ to lie on the outer
boundary~$\yb$~(so that~$\yub_\yim \yt^\yim=0$). This means that we
are focusing our attention on the foliation of the outer
boundary~$\yb$ into surfaces~$\yp_t$, rather than on the foliation of
the spacetime~$\ym$ into surfaces~$\ys_t$. This, in turn, implies
that we have to pass from the spacetime lapse~$\yns$ and
shift~$\yvs^\yis$ to the \emph{boundary} lapse~$\ynp$ and
shift~$\yvp^\yip$, and from the quantities~$\tr\ycps$, $\yus^\yim$,
$\yups^\yis$ to the quantities~$\tr\ycpb$, $\yupb^\yib$, $\yub^\yim$.

Let us now write the boundary contributions in the
Hamiltonian~(\ref{eq:ham1}) in terms of the new objects. Summing up
the following identities, which are obtained from
Eqs.~(\ref{eq:trps2}), (\ref{eq:moms-a}), (\ref{eq:norm0}),
and~(\ref{eq:norm1}):
\begin{align}
\begin{split}
\yns\fdil\tr\ycps=&\ep\yns\ypl\fdil\tr\ycpb +\ep\yns\ypl\ypz\fdil\tr\ycs\\ 
&\quad+\ep\yns\ypl\ypz \fdil\yups^\yim (\yub^\yjm \ydm_\yim \yupb_\yjm 
+ \ydm_\yim\ypa),
\end{split}\\
\begin{split}
2\yups_\yim \yvs_\yjm\smash{\frac{\yk}{\ygsd}}\yps^{\yim\yjm}=&
-\ep\yns\ypl\ypz\fdil\tr\ycs
+\ep\yns\ypl\ypz\lie_\yus\fdil\\
&\quad+\fdil\yvs^\yim (\yub^\yjm \ydm_\yim \yupb_\yjm 
+\ydm_\yim \ypa),
\end{split}\\
-\yns\lie_\yups\fdil=&-\ep\yns\ypl\lie_\yub\fdil 
-\ep\yns\ypl\ypz\lie_\yus\fdil,
\end{align}
and using the relations~$\ynp=\yns\ypl$ and~$\yvp^\yim =
\ep\yns\ypl\ypz\yups^\yim + \yvs^\yim$, we find that the boundary
integral in Eq.~(\ref{eq:ham1}) can be expressed as follows:
\begin{equation}
\begin{split}
\int_{\yp_t} (\yns\ye - \yvs^\yis \yj_\yis)=& 
\frac{1}{\yk}\int_{\yp_t}\ygpd[\ep\ynp(\fdil\tr\ycpb
- \lie_\yub\fdil)
+\fdil \yvp^\yim \yub^\yjm \ydm_\yim \yupb_\yjm \\
&\qquad+\fdil \yvp^\yim \ydb_\yim\ypa
- \ynp\ypa\ydb_\yim(\fdil\yupb^\yim)].
\end{split}
\end{equation}
Let us now consider the terms containing~$\ypa$, which can be
manipulated using the identities~$\ygbd=\ynp\ygpd$
and~$\yt^\yim=\ynp\yupb^\yim+\yvp^\yim$, to obtain:
\begin{multline}
\int_{\yp_t}\ygpd[\fdil \yvp^\yim \ydb_\yim\ypa
-\ynp\ypa\ydb_\yim(\fdil\yupb^\yim)]=\\
\begin{split}
&=\int_{\yp_t}\ygbd\biggl[\ydb_\yim
\biggl(\fdil\frac{\yvp^\yim}{\ynp}\ypa\biggr)
-\ypa\ydb_\yim\biggl(
\fdil\frac{\ynp\yupb^\yim +\yvp^\yim}{\ynp}\biggr)\biggr]\\
&=
\int_{\yp_t}\ygbd\ydb_\yim
\biggl(\fdil\frac{\yvp^\yim}{\ynp}\ypa\biggr)
-\int_{\yp_t} \ypa \lie_\yt(\fdil\ygpd).
\end{split}
\end{multline}
Note that the integral containing the total divergence can be
discarded, since, by Stokes' theorem, upon integration in time it
gives vanishing terms proportional to~$\yupb_\yim\yvp^\yim=0$
on~$\yp'$ and~$\yp''$. Moreover, it is easy to show that
\begin{equation}
\ypa \lie_\yt(\fdil\ygpd)\equiv
\ypp^{\yip\yjp} \lie_\yt\ygp_{\yip\yjp} 
+ \yppdil \lie_\yt\dil.
\end{equation}
Using the previous equations  we finally find
\begin{equation}
\begin{split}
\int_{\yp_t} (\yns\ye - \yvs^\yis \yj_\yis)=&
\frac{1}{\yk}\int_{\yp_t}\ygpd[
\ep\ynp(\fdil\tr\ycpb - \lie_\yub\fdil)
-\yvp^\yim \fdil\yupb^\yjm \ydm_\yim \yub_\yjm] \\
&-\int_{\yp_t}(
\ypp^{\yip\yjp} \lie_\yt\ygp_{\yip\yjp} + \yppdil \lie_\yt\dil).
\end{split}
\end{equation}
The last term can be discarded if we perform a Legendre
transformation different from~(\ref{eq:Leg1}):
\begin{equation}\label{eq:Leg2}
\int_t\yhii \!\equiv\!
\int_t\biggl[\int_{\ys_t}\!\! (\yps^{\yis\yjs} \lie_{\yt}\ygs_{\yis\yjs} 
\!+\! \ypsdil \lie_{\yt}\dil)
 \!+\! \int_{\yp_t}\!\! (\ypp^{\yip\yjp} \lie_\yt\ygp_{\yip\yjp} 
\!+\! \yppdil \lie_\yt\dil)\biggr] \!-\! \yli.
\end{equation}
In this way~$\ygp_{\yip\yjp}$,~$\dil|_\yp$,
and~$\ypp^{\yip\yjp}$,~$\yppdil$ are treated as canonical variables
and momenta on the same footing as~$\ygs_{\yis\yjs}$,~$\dil|_\ys$,
and~$\yps^{\yis\yjs}$,~$\ypsdil$.
We now have the new Hamiltonian:
\begin{equation}\label{eq:ham2}
\yhii =\int_{\ys_t} (\yns\yhd + \yvs^\yis \ymd_\yis)
+\int_{\yp_t} (\ynp\yei - \yvp^\yip \yji_\yip),
\end{equation}
where
\begin{subequations}\label{eq:ejP}
\begin{align}
\yei\defin &\frac{\ep}{\yk}\ygpd(\fdil\tr\ycpb - \lie_\yub\fdil),\\
\yji_\yis\defin& \frac{\ygpd}{\yk}\fdil\yupb^\yjm \ydm_\yim \yub_\yjm.
\end{align}
\end{subequations}

This Hamiltonian reduces, in the non-dilatonic case and when~$\yb$ is
timelike, to Booth and Mann's Hamiltonian. Anologously to the BM
Hamiltonian, $\yhii$~has no explicit dependence upon the intersection
angle between~$\yb$ and~$\ys_t$. This happens because all quantities
in the boundary term of~(\ref{eq:ham2}) are defined considering a
local, natural spacetime foliation of~$\ym$ into slices~$\ysp_t$
orthogonal to the outer boundary.

It is interesting to compare the boundary term of~$\yhii$,
Eq.~(\ref{eq:ham2}), with that of~$\yhi$, Eq.~(\ref{eq:ham1}); this
can simply be done by rewriting~$\ynp$, $\yvp^\yip$, $\tr\ycpb$,
$\yupb^\yib$, and~$\yub^\yim$ in terms of~$\yns$, $\yvs^\yis$,
$\tr\ycps$, $\yus^\yim$, $\yups^\yis$. The result is:
\begin{multline}\label{eq:compare}
\int_{\yp_t} (\ynp\yei - \yvp^\yip \yji_\yip)=\\
\begin{split}
&=\frac{1}{\yk}\int_{\yp_t} \ygpd 
\biggl[\yns\fdil\tr\ycps - \yns\lie_\yups\fdil 
- \fdil\yvs^\yim{\ygp_\yim}^\yjm\yds_\yjm\ypa
+ 2\yk\yvs_\yim\yups_\yjm \frac{\yps^{\yim\yjm}}{\ygsd}\biggr]\\
&=\frac{1}{\yk}\int_{\yp_t} \ygpd 
\biggl[\yns\fdil\tr\ycps - \yns\lie_\yups\fdil 
+ \ypa\, {\ygp_\yim}^\yjm\yds_\yjm(\fdil{\ygp^\yim}_\ykm\yvs^\ykm)
+ 2\yk\yvs_\yim\yups_\yjm \frac{\yps^{\yim\yjm}}{\ygsd}\biggr].\\
\end{split}
\end{multline}

\subsubsection{Background terms}

It is a well-known fact that we can subtract from the gravitational
action (and thus from the Hamiltonian) a reference term~$\ylio$,
which has to be a functional of the boundary metric only, without
affecting the equations of motion of the system. Subtracting such a
term corresponds to redefining the zero-point energy and
momentum~\cite{BY}. This may be necessary when we want to renormalize
divergent quantities, which may appear in the Hamiltonian when we
consider an outer boundary at infinity. Usually one chooses this term
in order to have vanishing energy and momentum for a given reference
spacetime (e.g.~Minkowski or anti-de~Sitter spacetime).

For the Hamiltonian $\yhi$ of Eq.~(\ref{eq:ham1}), the reference term
can be defined in the following way (note that another equivalent,
but local, definition can be given along the lines of
ref.~\cite[Sect.~III~D]{BM1}). First we embed the boundary~$(\yb,
\ygb_{\yib\yjb}, \dil|_\yb)$ into the reference spacetime~$(\ymo,
\ygmo_{\yim\yjm}, \dilo)$ in such a way that the metric and dilaton
induced on~$\yb$ by the embedding agree with those induced from~$\ym$
(we may call it an isometric and `isodilatonical' embedding).
Moreover we must require that~$\ymo$ be foliated in such a way
that~$\ypz$~(see~Eq.~(\ref{eq:beta})) has the same value in~$\ym$ and
in~$\ymo$.

These conditions together imply that~$\ygpd \yns$, $\ypl$, $\ypa$,
$\ydb_\yib$, $\yupb^\yib$, and~$\fdil|_\yb$ are the same in the two
spacetimes. Hence, we define the reference term to be
\begin{equation}
\yhio_{\yp_t}\defin-\int_{\yp_t} (\yns\ye - \yvs^\yis\yj_\yis)\quad
\text{calculated with respect to~$\ymo$}.
\end{equation}
With this definition the boundary term~$\yhi_{\yp_t}$ becomes
explicitly:
\begin{equation}
\begin{split}
\yhi_{\yp_t}=&\frac{1}{\yk}\int_{\yp_t}\ygpd\biggl
\{\yns[\fdil(\tr\ycps - \tr\ycpso)
-(\lie_\yups\fdil - \lie_\yupso\fdilo)]\\
&\qquad- \yk\biggl(\yvs_\yis \yups_\yjs\frac{\yps^{\yis\yjs}}{\ygsd}
-\yvso_\yis \yupso_\yjs\frac{\ypso^{\yis\yjs}}{\ygsdo}\biggr)
\biggr\},
\end{split}
\end{equation}
where all objects with an under-bar are evolved on the reference
spacetime~$\ymo$. Note that the term containing the explicit
dependence on the hyperbolic angle~$\ypa$ has disappeared, for it is
the same in both spacetimes: this makes the presence of the reference
term a necessary feature in the case of the Hamiltonian~$\yhi$ of
Eq.~(\ref{eq:ham1}).

The situation is different in the case of the Hamiltonian~$\yhii$ of
Eq.~(\ref{eq:ham2}). We still require an isometric and isodilatonical
embedding of~$\yb$ in the reference spacetime~$\ymo$, but now we do
not impose any requirement about the foliation of~$\ymo$ and the
intersection angle; yet this weaker condition implies that the
\emph{boundary} lapse and shift agree in both spacetimes. The
reference term is then defined by
\begin{equation}
\begin{split}
\yhiio_{\yp_t}\defin&-\int_{\yp_t} (\ynp\yei - \yvp^\yip\yji_\yip)\quad
\text{calculated with respect to~$\ymo$} \\
\equiv&-\int_{\yp_t} (\ynp\yeio - \yvp^\yip\yjio_\yip),
\end{split}
\end{equation}
where
\begin{subequations}
\begin{align}
\yeio\defin &\frac{\ep}{\yk}\ygpd(\fdil\tr\ycpbo 
-\lie_\yubo\fdilo),\\
\yjio_\yis\defin& \frac{1}{\yk} \ygpd\fdil\yupb^\yjm \ydmo_\yim \yubo_\yjm;
\end{align}
\end{subequations}
The boundary term in~$\yhii$ becomes now
\begin{equation}
\begin{split}
\yhii_{\yp_t}&=\int_{\yp_t} [\ynp(\yei-\yeio) 
- \yvp^\yip(\yji_\yip -\yjio_\yip)]\\
&=\frac{1}{\yk}\int_{\yp_t}\ygpd\{\ep\ynp[\fdil(\tr\ycpb-\tr\ycpbo) 
- (\lie_\yub\fdil-\lie_\yubo\fdilo)]\\
&\qquad+\fdil\yvp^\yim\yupb^\yjm (\ydm_\yim \yub_\yjm 
-\ydmo_\yim \yubo_\yjm)\}.
\end{split}
\end{equation}

In this case the reference term is not necessary to
eliminate explicit angle dependence in the Hamiltonian, for~$\yhii$
has none by construction.

\subsubsection{Null and non-transversal outer boundaries}

All the results obtained so far hold, generally, for a spacetime with
a timelike or transversal spacelike outer boundary. The formalism
developed in the previous subsections can deal with these two cases
but is not meant to deal with a null or non-transversal spacelike
outer boundary. In the first case, the main reason is that the
action~(\ref{eq:lag}) becomes ill-behaved whenever one considers the
limit of a null~$\yb$, for the integrands in the $\yb$-~and
$\yp$-surface integrals diverge; one should define a new action with
appropriate boundary terms before going on to derive Hamiltonians. In
the second case, the foliation is no more well-defined.

Yet, no one prevents us from considering a null~$\yb$ or a
non-transversal spacelike~$\yb$ as limits of a regular outer
boundary. We are forced to allow for these possibilities if we want
to consider e.g.~lightlike generators or generators of spatial
translations.

The null-$\yb$ limit corresponds to~$\ypa\rightarrow\pm\infty$.
Eqs.~(\ref{eq:ejS}) and (\ref{eq:depangle}) show that the boundary
term of the first Hamiltonian~$\yhi$ diverges in this limit; this
divergence can be easily cured: it disappears upon subtracting
from~$\yhi$ a reference term as we have discussed in the previous
subsection. The boundary term of the second Hamiltonian~$\yhii$
diverges as well in general, as one can see from Eqs.~(\ref{eq:ejP})
and~(\ref{eq:compare}); however, this is not always true and we see
that~$\yhii$ is well-behaved, for~$\ypa\rightarrow\pm\infty$, when
the component of~$\yt^\yim$ tangent to~$\yp_t$,
namely~${\ygp^\yim}_\yjm \yt^\yjm = {\ygp^\yim}_\yjm \yvs^\yjm =
\yvp^\yim$, is divergence-free on~$\yp_t$,
i.e.~when~${\ygp^\yim}_\yjm \yds_\yim ({\ygp^\yim}_\yjm \yvs^\yim) =
0$; note that this is trivially verified in a two-dimensional
spacetime ($\yd=1$), since~$\yp_t$ is a point (or a finite collection
of points) in that case.

The non-transversal-$\yb$ limit corresponds, instead,
to~$\yns,\ypa\rightarrow 0$. In this case it is easy to verify, again
by means of Eqs.~(\ref{eq:ejS}), (\ref{eq:depangle}),
and~(\ref{eq:ejP}), (\ref{eq:compare}), that the surface terms of
both Hamiltonians are well-behaved and yield the same limit, which,
for constant shift and in asymptotically flat or anti~de-Sitter
spacetimes, corresponds to the total momentum
(cf.~e.g.~refs.~\cite{RT,BH} for the non-dilatonic case):
\begin{align}\label{eq:lim}
\int_{\yp_t} (\yns\ye - \yvs^\yis \yj_\yis)
&\longrightarrow 2\int_{\yp_t}\ygpd \yvs_\yim \yups_\yjm
\frac{\yps^{\yim\yjm}}{\ygsd},\\
\int_{\yp_t} (\ynp\yei - \yvp^\yip \yji_\yip)
&\longrightarrow 2\int_{\yp_t}\ygpd \yvs_\yim \yups_\yjm
\frac{\yps^{\yim\yjm}}{\ygsd}.
\end{align}

\section{Discussion}

In this paper we have derived two Hamiltonians for a general
$(\yd\!+\!1)$-di\-men\-sion\-al dilaton gravity
theory,~$\yhi$~(Eq.~(\ref{eq:ham1}))
and~$\yhii$~(Eq.~(\ref{eq:ham2})), which generalize Hawking and
Hunter's~\cite{HH} and Booth and Mann's~\cite{BM1} Hamiltonian
respectively. For the purposes of the present discussion we can
call~$\yhi$ `bulk-oriented Hamiltonian' and~$\yhii$
`boundary-oriented Hamiltonian'.

When we use the bulk-oriented Hamiltonian, we focus our attention
mainly on the foliation of~$\ym$ into spacelike hypersurfaces and we
use the `bulk' lapse~$\yns$ and shift~$\yvs^\yis$, together with
other~$\ys_t$-related objects ($\tr\ycps$, $\lie_\yus \dil$, etc.).
Conversely, when we use the boundary-oriented Hamiltonian we assume
that the initial surface~$\yp'$ is time-evolved along~$\yb$, we
restrict our attention to the boundary foliation and we use the
boundary lapse~$\ynp$ and shift~$\yvp^\yip$, and other boundary
objects~($\tr\ycpb$, $\lie_\yub \dil$, etc.).\footnote{The
bulk-oriented approach is used e.g.~by Lau~\cite{Lau1,Lau2}, Hawking
and Horowitz~\cite{HaHo}, Hawking and Hunter~\cite{HH}, and
implicitly also by DeWitt~\cite{dW}, Regge and Teitelboim~\cite{RT},
Brown and Henneaux~\cite{BH}, Cadoni and Mignemi~\cite{CaMi,CaMi2},
et al. The boundary-oriented approach is followed by Brown and
York~\cite{BY}, Bose and Dadhich~\cite{BD}, Kijowski~\cite{Ki},
Brown, Creighton and Mann~\cite{BCM}, Creighton and Mann~\cite{CM},
Booth and Mann~\cite{BM1,BM2}, Brown, Lau, and York~\cite{BLY}, et
al.}

We have seen also that one Hamiltonian has merits where the other has
drawbacks, and vice versa. The bulk-oriented one allows to consider
all kinds of generators (spacelike, timelike, null), but contains an
explicit dependence upon the hyperbolic angle of the foliation
$\alpha$---which diverges in the limit of a null outer
boundary---needing an additive spacetime reference term. Conversely,
the boundary-oriented Hamiltonian has no explicit dependence on the
intersection between the slices and the outer boundary, yet forces us
to modify the latter (considering even the possibility of a spacelike
case), and in some cases to resort to limit procedures, in order to
study a generic evolution generator.

Apart from the fact that the two Hamiltonians correspond to a
different choice of thermodynamical ensembles (see
e.g.~Kijowski~\cite{Ki}), it is evident that they are useful in two
complementary situations. $\yhi$~is the natural choice for the
Hamiltonian when one is dealing with a spatially non-compact
spacetime, whereas~$\yhii$ is useful in the case of a bounded
spacetime.

In a spatially non-compact spacetime we may want to consider
e.g.~spatial translations, which usually belong to the group of
automorphisms of the manifold. In this case the bulk-oriented
Hamiltonian allows us to study the generators of translations. We
first introduce a boundary at finite distance, study the generators
on this boundary, then push the boundary to infinity to study the
asymptotic behavior of our generators. The fact that the generators
map the manifold out of the boundary is of no importance, since the
boundary is introduced only to be pushed to infinity. Moreover, in
this case the explicit dependence upon the angle foliation is not
problematic: a spatially non-compact spacetime usually needs a
reference spacetime for renormalizing possible divergences, and the
dependence on the angle~$\alpha$ can be eliminated together with the
divergences by subtracting from~$\yhi$ a reference term.
 
On the other hand, when we deal with a spatially bounded manifold,
usually, we do not consider transformations like e.g.~spatial
translations, for they are not automorphisms of the spacetime. In
this case one naturally uses the boundary-oriented
Hamiltonian~$\yhii$ , which has no dependence on~$\alpha$. Moreover,
the formalism we have developed in this paper enables one to
use~$\yhii$ in spacetimes with almost all kind of boundaries, hence
makes~$\yhii$ as much versatile as~$\yhi$ for the study of almost all
kinds of generators.

\end{document}